\documentstyle[twocolumn,psfig,rotating]{mn}
\def\pmb#1{\setbox0=\hbox{#1}%
\kern-.025em\copy0\kern-\wd0
\kern.05em\copy0\kern-\wd0
\kern-.025em\raise.0433em\box0}

\begin{document}
\title[Element segregation]
{Element segregation in giant galaxies and X-ray clusters}

\author[Chuzhoy \& Loeb ]{Leonid Chuzhoy
and
Abraham Loeb\\
Physics Department,
Technion, Haifa 32000, Israel\\
Astronomy Department, Harvard University, 60 Garden Street, Cambridge, 
MA 02138, USA\\
E-mail: cleonid@tx.technion.ac.il,  aloeb@cfa.harvard.edu
}

\maketitle

\begin{abstract}
We examine the process of element segregation by gravity in giant
elliptical galaxies and X-ray clusters. We solve the full set of flow
equations developed by Burgers for a multi-component fluid, under the
assumption that magnetic fields slow the diffusion by a constant factor
$F_B$. Compared to the previous calculations that neglected the residual
heat flow terms, we find that diffusion is faster by $\sim 20\%$.  In
clusters we find that the diffusion changes the local abundance typically
by factors of $1+0.3(T/10^8~{\rm K})^{1.5}/F_B$ and $1+0.15(T/10^8~{\rm
K})^{1.5}/F_B$ for helium and heavy elements, respectively, where $T$ is
the gas temperature. In elliptical galaxies, the corresponding factors are
$1+0.2(T/10^7~{\rm K})^{1.5}/F_B$ and $1+0.1(T/10^7~{\rm K})^{1.5}/F_B$,
respectively. If the suppression factor $F_B$ is modest, diffusion could
significantly affect observational properties of hot X-ray clusters and cD
galaxies. In particular, diffusion steepens the baryon distribution,
increases the total X-ray luminosity, and changes the spectrum and
evolution of stars that form out of the helium-rich gas.  Detection of
these diffusion signatures would allow to gauge the significance of the
magnetic fields, which also inhibit thermal heat conduction as a mechanism
for eliminating cooling flows in the same environments.
\end{abstract}

\begin{keywords}
galaxies: clusters: general -- cosmology:theory -- dark matter -- X-rays: galaxies.
\end{keywords}

\section{Introduction}
In a multicomponent plasma, gravity acts so as to separate different
elements. The thermal equilibrium state for the number density of particles
of mass $m_{\rm i}$ and temperature $T$ is proportional to the Boltzmann factor,
$e^{-m_{\rm i}\phi(r)/kT}$, where $\phi(r)$ is the gravitational potential. Thus
in self-gravitating objects formed out of a homogeneous mixture of
elements, the heavier elements (such as helium, metals) sink towards the
center and the lighter elements (such as hydrogen) rise up. Fabian \&
Pringle (1977) estimated that in high-temperature X-ray clusters
significant abundance gradients might arise for metals. However, their
calculation did not allow for the suppression of diffusion by magnetic
fields, which could play an important role in clusters. Since a charged
particle is confined to move along magnetic field lines, diffusion can
proceed uninhibited only on the coherence length-scale of the magnetic
field. In clusters the scale of magnetic field is likely to be small and
the diffusion speed could be suppressed by several orders of magnitude
(Chandran \& Cowley 1998), making its effect negligible. In recent years,
however, it has been suggested that turbulence may suppress transport
processes by only a modest factor of a few (Narayan \& Medvedev 2001;
Malyshkin 2001). Such models are motivated by the lack of observational
evidence for cooling flows in the cores of clusters and elliptical galaxies
(Mollendi \& Pizzolato 2001; Fabian et al. 2001). Cooling flows would be
suppressed if thermal heat conduction by electrons is not strongly
inhibited by the magnetic fields.

In this paper we estimate the impact of diffusion on observational
properties of galaxies and X-ray clusters for different magnetic
suppression factors. Unlike previous analyses (Fabian \& Pringle 1977,
Rephaeli 1978, Gilfanov \& Syunyaev 1984, Qin \& Wu 2000, Chuzhoy \& Nusser
2003), we solve the full diffusion equations derived by Burgers (1969).  We
find that if the suppression factor is modest, then diffusion would lead to
significant abundance gradients for all elements. In extreme cases, the
abundance of helium (which except for deuterium is the fastest diffusing
element) can grow above that of hydrogen. Because helium sedimentation
changes the mean molecular weight, the gas density profile steepens and the
total luminosity of the cluster increases.  Since the diffusion speed
depends strongly on temperature, this effect may contribute to the
luminosity-temperature dependence observed in clusters (Chuzhoy \& Nusser
2003). The change in the mean molecular weight would also affect mass
estimates of hot clusters and elliptical galaxies under the assumption of
hydrostatic equilibrium (Qin \& Wu 2000, Chuzhoy \& Nusser 2003).  In
addition to its effect on the X-ray emitting gas, the increase in the
helium abundance could affect the spectrum and evolution of stars that form
out of this gas.

Because the diffusion speed in a plasma scales with temperature as
$T^{3/2}$, all previous work was restricted to X-ray clusters which are the
hottest virialized objects. However, we find that because of the lower gas mass fraction,
diffusion would proceed faster in hot cD galaxies with $T\sim 2\times 10^7$
K than in clusters with $T<10^8$ K.

The paper is organized as follows. In \S 2.1 we write the general diffusion
equations. In \S 2.2 we evaluate the diffusion timescales and describe the
diffusion in the perturbative regime, where the fractional deviations of the
abundances from uniformity are small. In \S 3 we describe diffusion
in the non-linear regime, assuming the NFW mass profile for the dark matter
(Navarro, Frenk \& White 1997). We summarize our results in \S 4.

\section{Diffusion equations}
\subsection{General equations}

Each species of particles {\it s} is described by a distribution function
$F_{\rm s}(x,v,t)$ normalized to unit integral, a mean number density $n_{\rm s}$, an
ionic charge $q_{\rm s}=Z_{\rm s} e$ and a particle mass $m_{\rm s}$. The cross-section for
Coulomb scattering between particles of species $s$ and of species $t$ is
given by
\begin{equation}
\sigma_{\rm st}=2\sqrt{\pi}e^4Z_{\rm s}^2Z_{\rm t}^2(k_BT)^{-2}\ln{\Lambda_{\rm st}},
\end{equation}
where $\ln{\Lambda_{\rm st}}$ is the Coulomb logarithm. For simplicity, we
approximate the Coulomb logarithm for all species by a constant,
$\ln{\Lambda_{\rm st}}=40$, which is characteristic for clusters. The friction
coefficient between species ${\it s}$ and ${\it t}$ is
\begin{equation}
K_{\rm st}=(2/3)n_{\rm s} n_{\rm t} \sigma_{\rm st}\left(\frac{2k_{\rm B} T m_{\rm s}
m_{\rm t}}{m_{\rm s}+m_{\rm t}}\right)^{1/2}.
\end{equation}
The mean fluid velocity is
\begin{equation}
{\bf u}=(\mathop{\Sigma}_{\rm s} n_{\rm s} m_{\rm s} {\bf u_s})/(\mathop{\Sigma}_{\rm s} n_{\rm s}
m_{\rm s}),
\end{equation}
where ${\bf u_{\rm s}}$ is the mean fluid velocity of each species. The diffusion
velocity of species ${\it s}$ is defined as
\begin{equation}
{\bf w_{\rm s}}={\bf u_{\rm s}}-{\bf u}.
\end{equation}
and its ``residual heat flow vector'' is (Burgers 1969)
\begin{equation}
{\bf r_{\rm s}}=\frac{m_{\rm s}}{2k_{\rm B} T}\int F_{\rm s}({\bf v}-{\bf u})|{\bf v}-{\bf u}|^2d{\bf
v}-\frac{5}{2}{\bf w_{\rm s}}.
\end{equation}
If the gas is in hydrostatic equilibrium (i.e. ${\bf \nabla} p/\rho={\bf
g}$), the Burgers equations for mass, momentum and energy conservation in
spherical symmetry, as formulated by Thoul, Bahcall \& Loeb (1994), are
\begin{eqnarray}
\label{cont2}
\frac{\partial n_{\rm s}}{\partial t}=-\frac{1}{r^2}\frac{\partial (r^2 n_{\rm s}
u_{\rm s})}{\partial r}, \hspace{42mm} \\
\label{burg1}
\frac{d(n_{\rm s}k_B T)}{dr}+n_{\rm s} m_{\rm s} g-n_{\rm s} Z_{\rm s} eE=\hspace{28mm}\nonumber\\
\mathop{\Sigma}_{\rm t} K_{\rm st}[(w_{\rm t}-w_{\rm s})+0.6(x_{\rm st}r_{\rm s}-y_{\rm st}r_{\rm t})], \\
\label{burg2}
\frac{5}{2}n_{\rm s} k_B\frac{dT}{dr}= -0.8K_{ss}r_{\rm s}+ \mathop{\Sigma}_{\rm t}
K_{\rm st}\{\frac{3}{2}x_{\rm st}(w_{\rm s}-w_{\rm t})\nonumber\\-y_{\rm st}[1.6x_{\rm st}(r_{\rm s}+r_{\rm t})+Y_{\rm st}r_{\rm s}-4.3x_{\rm st}r_{\rm t}]\},
\end{eqnarray}
where $g$ is the gravitational acceleration, $E$ is the electric field,
$x_{\rm st}=m_{\rm t}/(m_{\rm s}+m_{\rm t})$, $y_{\rm st}=m_{\rm s}/(m_{\rm s}+m_{\rm t})$ and
$Y_{\rm st}=3y_{\rm st}+1.3x_{\rm st}m_{\rm t}/m_{\rm s}$.  From equations (\ref{burg1}) and
(\ref{burg2}) we can obtain a solution for the diffusion velocity and the
heat flow of each species
\begin{eqnarray}
\label{vel}
w_{\rm s}=\frac{T^{5/2}}{n_{\rm p}}\left(\mathop{\Sigma}_i I_i^s\frac{\partial\ln
n_i}{\partial r}+I_T^s\frac{\partial \ln T}{\partial r}\right) ,\\
\label{heat1}
r_{\rm s}=\frac{T^{5/2}}{n_{\rm p}}\left(\mathop{\Sigma}_i J_i^s\frac{\partial\ln
n_i}{\partial r}+J_T^s\frac{\partial \ln T}{\partial r}\right) ,
\end{eqnarray}
where the coefficients $I_i^s$, $I_T^s$, $J_i^s$ and $J_T^s$ depend only on
the composition of the gas.
In the presence of magnetic fields, we introduce a constant suppression
factor, $F_B^{-1}$, for all these coefficients. The actual suppression
factor in reality should depend on the geometry and fluctuations of the
magnetic field which may vary with position; but the lack of detailed data
on these properties calls for a simple-minded approach of the type we
adopt.

If the diffusion is ``switched-on'' when the gas is in hydrostatic
equilibrium, then initially there would be no mass flow ($u=0$). For each
sinking helium ion there must be approximately four protons and two
electrons that rise up. This means that there is a net outflow of
particles, which leads to a decrease in the total pressure ($p=\Sigma n_{\rm s}
k_B T$). For a fixed mass density, the drop in gas pressure means that the
assumption of hydrostatic equilibrium (${\bf \nabla} p/\rho=g$) can
not hold once diffusion starts to operate. In practice the sound speed of
the gas is much larger than the diffusion speed, i.e. the dynamical
timescale of the system is much shorter than the diffusion timescale. Thus,
the gas will always be close to hydrostatic equilibrium and the above
equations for the relative velocities of all species remain valid. 
To find a small mean fluid velocity, we solve the equation of motion, 
\begin{equation}
\label{newt}
\frac{du}{dt}=-{\nabla p \over \rho} +g ,
\end{equation}
where $g=g(r)$ is the gravitational field dictated by the total mass
distribution.

 If we wish to evaluate the diffusion velocity of hydrogen
($s\equiv p$) and helium ($s\equiv \alpha$), which are the dominant
components of the cosmic plasma, we may neglect the contribution of all
other species. Thus, the drift velocity of helium ions is
\begin{equation}
\label{helvel}
w_\alpha=\frac{T^{5/2}}{n_{\rm p}}\left( I_{\rm p}^\alpha\frac{\partial\ln
n_{\rm p}}{\partial r}+I_\alpha^\alpha\frac{\partial\ln n_\alpha}{\partial
r}+I_T^\alpha\frac{\partial \ln T}{\partial r}\right),
\end{equation}
where $I_{\rm p}$, $I_\alpha$ and $I_T$ are functions of hydrogen mass fraction
$X$. The velocity of the protons can be obtained simply from the relation
$w_{\rm p}=w_\alpha(X-1)/X$.

The diffusion velocity of heavy elements is determined primarily by the
drag forces from hydrogen and helium (Chuzhoy \& Nusser 2003). Helium
provides an inward radial friction force for the metals, while protons push
them radially outwards.  The resulting velocity can be determined by
balancing these friction forces. Together with the pressure gradient force,
the gravitational and electric forces for the metals increase their
diffusion velocity relative to hydrogen by a factor $\sim (1+4/Z)$ (Chuzhoy
\& Nusser 2003). For heavy elements such as iron ($Z=26$) this correction
can be ignored.  Using equation (\ref{burg1}) we obtain the approximation
for the diffusion velocity of all heavy elements
\begin{equation}
\label{met}
w_z=\frac{(1-X)w_\alpha-1.2(1-X)r_\alpha-0.6X r_{\rm p}}{(2-X)}
\end{equation}

The total heat flow and the temperature evolution are given by 
\begin{eqnarray}
Q&=&\mathop{\Sigma}_{\rm s} (n_{\rm s} r_{\rm s} kT), \\
\mathop{\Sigma}_{\rm s}\frac{\partial (\frac{3}{2}n_{\rm s} kT)}{\partial t}&=&\frac{1}{r^2}\frac{\partial (r^2 Q)}{\partial r}
\end{eqnarray}
Somewhat surprisingly it turns out that $Q$ and $dT/dt$ are generally
nonzero even in the absence of a temperature gradient.  As implied by
equation (\ref{heat1}), the density gradients can alone be responsible for
the heat transport and eventually cause temperature gradients.  However,
because the coefficients $J_{\rm s}$ are several orders of magnitude
smaller than $J_T$, the effect of the heat flow on the temperature profile
is negligible in absence of an initial temperature gradient. We also find from Burgers equations that the density gradients can not produce the heat transport and temperature gradients after all elements reach equilibrium distribution.

\subsection{Diffusion in the perturbative regime}

The helium to hydrogen ratio starts from the uniform distribution set by
the big bang. In this section, we derive simple analytic results for small
deviations from this uniform initial state.  In the interiors of X-ray
clusters and elliptical galaxies, the density changes by a much larger
factor than the temperature and so $\vert \frac{\partial\ln n}{\partial
r}\vert \gg \vert \frac{\partial \ln T}{\partial r}\vert$. Since $I_{\rm s}$ and
$I_T$ are of similar magnitude, equation (\ref{helvel}) simplifies to
\begin{equation}
\label{helvel0}
w_{\alpha,0}=\frac{T^{5/2}}{n_{\rm p}}(I_{\rm p}^\alpha+I_\alpha^\alpha)\frac{\partial\ln
n_{\rm p}}{\partial r}
\end{equation}
Initially the hydrogen fraction $X\approx 3/4$ everywhere and
$I=I_{\rm p}^\alpha+I_\alpha^\alpha=2.6\times 10^{7}/F_B \; {\rm cm^{-1}
s^{-1}K^{-5/2}}$, where $F_B$ is the suppression factor due to the magnetic
field. Using the assumption of approximate hydrostatic equilibrium, we can
rewrite equation (\ref{helvel0}) as
\begin{eqnarray}
\label{helvel1}
w_{\alpha,0}=60\; {\rm km\; s^{-1}} \hspace{40mm}\\ \nonumber
\times\left(\frac{g}{10^{-7.5}\; {\rm cm\;s^{-2}}}\right)\left(\frac{T}{10^8
{\rm K}}\right)^{3/2}\left(\frac{n_{\rm p}}{10^{-3}\; {\rm cm}^{-3}}\right)^{-1}F_B^{-1}
\end{eqnarray}
The velocity obtained above is $\sim 20\%$ higher than the estimate by Chuzhoy \&
Nusser (2003) that neglected the heat flow terms and $\sim30\%$ lower than
the estimate by Qin \& Wu (2000) that also neglected electric fields.

Substituting equation (\ref{helvel1}) into the continuity equation gives
\begin{equation}
\label{cont}
\frac{\partial n_\alpha}{\partial t}=\frac{n_\alpha}{12\; {\rm Gyr}}
\left(\frac{f_{g}}{0.1}\right)^{-1}\left(\frac{T}{10^8 {\rm
K}}\right)^{3/2}F_B^{-1}
\end{equation}
where $f_{g}$ is the local gas fraction. Equation (\ref{cont}) implies that at
the initial stages of diffusion, the local helium density grows as
$e^{t/\tau}\sim(1+t/\tau)$, where
\begin{equation}
\label{tau}
\tau=12\;{\rm Gyr}\;\left(\frac{f_{g}}{0.1}\right)\left(\frac{T}{10^8
{\rm K}}\right)^{-3/2}F_B.
\end{equation}
Similarly the density of hydrogen declines as $e^{-t/3\tau}$ for $X=3/4$,
so that the relative abundance of helium grows as $e^{4t/3\tau}$. The
initial diffusion velocity of metals is $\sim {1\over 3}$ of helium
velocity, and so the metallicity grows as $\sim e^{2t/3\tau}$.

\section{Diffusion in galaxies and clusters}
For the radial distribution of the mass density of the dark matter, we
assume the NFW profile (Navarro, Frenk \& White 1997)
\begin{equation}
\label{grav}
\rho_d(r)=\frac{\rho_{\rm s}}{(cr/r_{vir})(1+cr/r_{vir})^2},
\end{equation}
where $\rho_{\rm s}=const$, $c$ is the concentration parameter and $r_{vir}$ is
the virial radius. We make separate calculations for $c=4$ and $c=15$,
which are characteristic values for clusters and elliptical galaxies,
respectively (Navarro, Frenk \& White 1997; Wechsler et al. 2002).  We
assume that the gas is initially isothermal and in hydrostatic
equilibrium. Neglecting the contribution of the baryons to the
gravitational potential, one obtains the following radial mass profile for
the gas (Makino, Sasaki \& Suto 1998)
\begin{equation}
\label{mak}
\rho_{\rm g}(r)=\rho_0 e^{-\eta}(1+cr/r_{vir})^{\eta r_{vir}/cr},
\end{equation}
where $\eta=4\pi G\mu m_{\rm p}\rho_{\rm s} r_{vir}^2/c^2 kT$.  Assuming that the gas
is at the virial temperature, $\eta=10$ and $\eta=16.5$ correspond to $c=4$
and $c=15$ respectively.  For clusters, we chose the ratio $\rho_0/\rho_{\rm s}$
so that the gas mass fraction within the virial radius is equal to the
cosmic value of $(\Omega_b/\Omega_m)=0.17$ inferred by WMAP (Bennett et
al. 2003).  In elliptical galaxies, typically $\sim 90\%$ of the baryonic
mass is in stars (Brigenti \& Mathews 1997), and so we take the gas
fraction ten times lower.  We assume that the initial abundances are
uniform and that the temperature remains constant in time as well as in
space.

\begin{figure} 
\centering \mbox{\psfig{figure=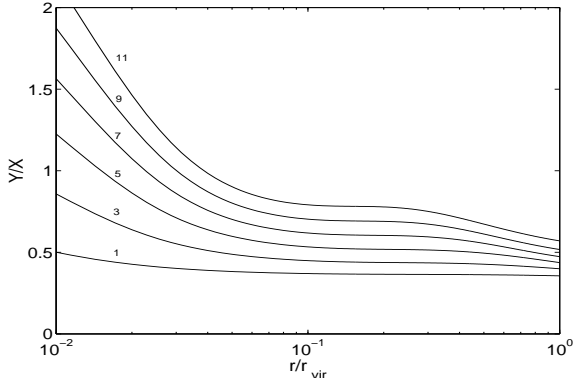,height=2.0in,width=3.0in}}
\caption{Helium to hydrogen mass density ratio vs radius in clusters. The
numbers on the curves correspond to the elapsed time in Gyr divided by
$F_B\times (T/10^8 K)^{-3/2}$. }
\label{fig1}
\end{figure}

\begin{figure} 
\centering
\mbox{\psfig{figure=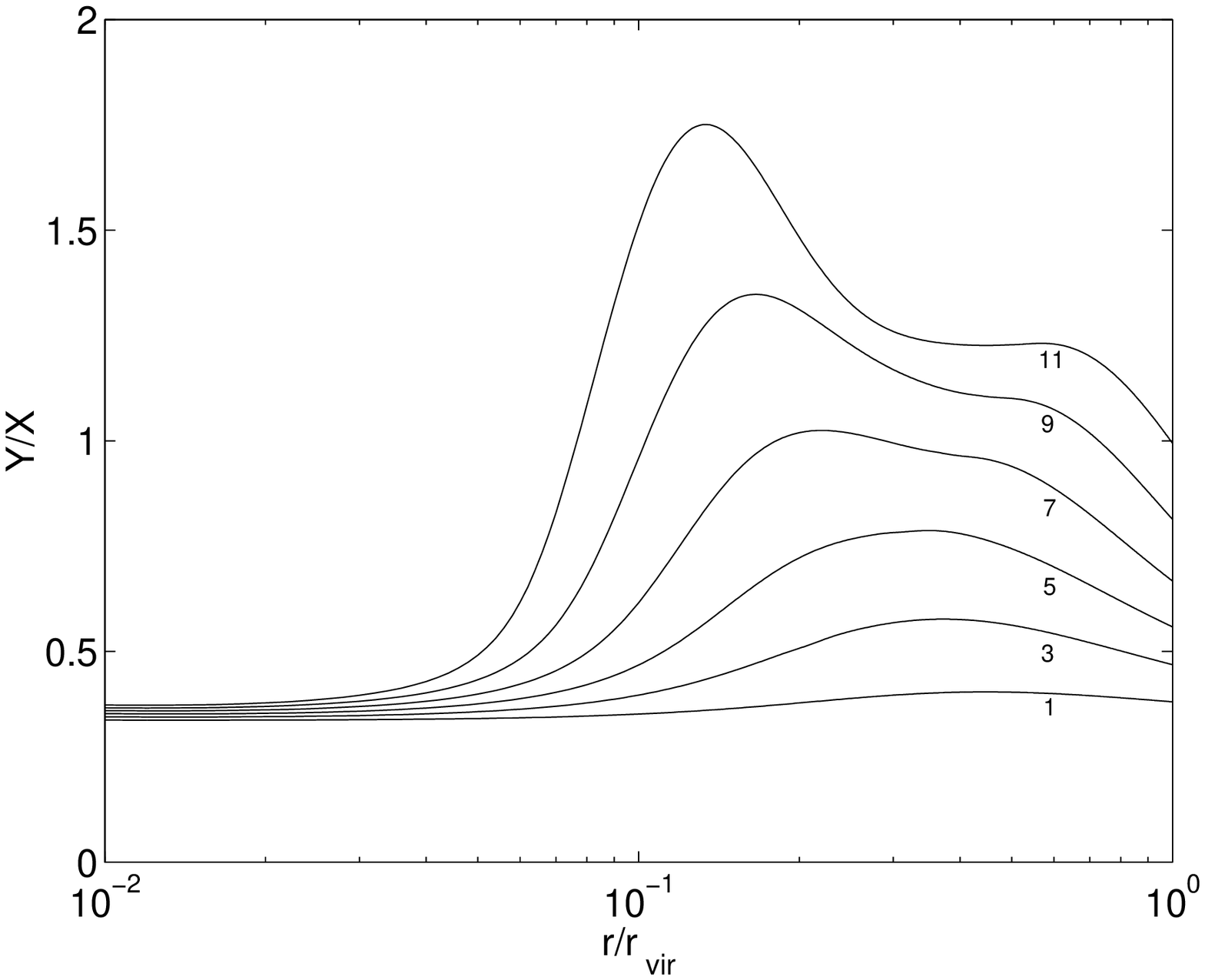,height=2.0in,width=3.0in}}
\caption{Same as Fig. \ref{fig1} for elliptical galaxies. }
\label{fig2}
\end{figure}

\begin{figure} 
\centering
\mbox{\psfig{figure=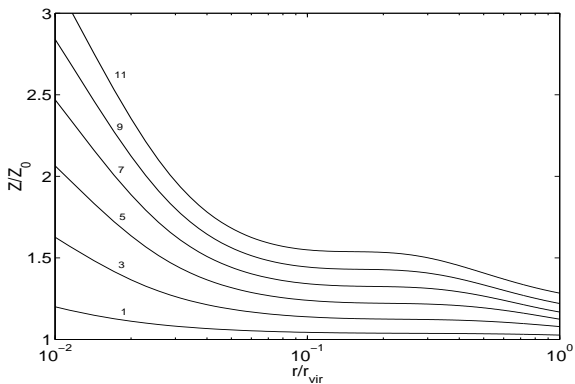,height=2.0in,width=3.0in}}
\caption{ Metallicity vs radius in clusters. The plot is normalized by the initial metallicity $Z_0$. The numbers on the curves correspond to the elapsed time in Gyr divided by $F_B\times (T/10^8 K)^{-3/2}$. }
\label{fig3}
\end{figure}

\begin{figure} 
\centering
\mbox{\psfig{figure=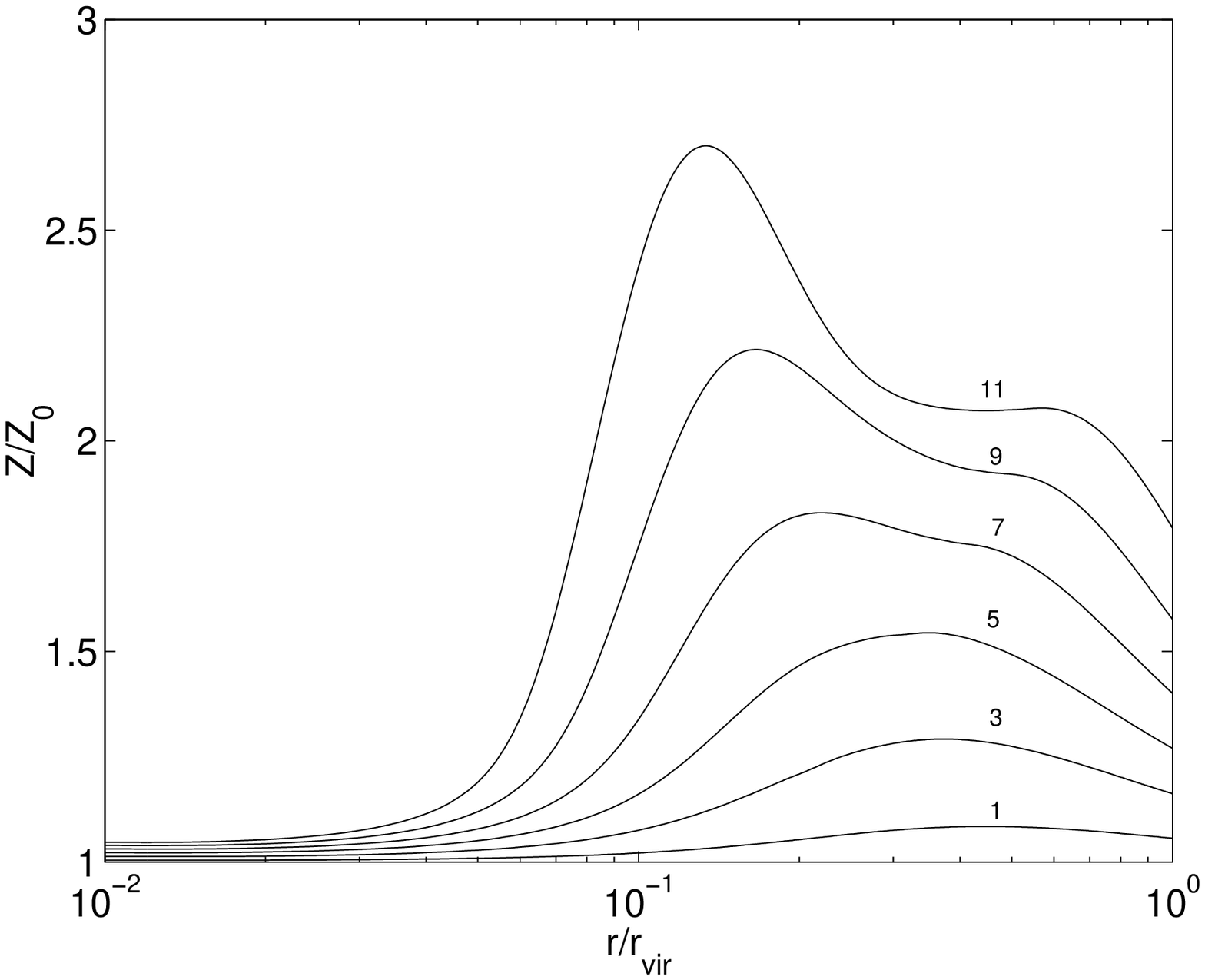,height=2.0in,width=3.0in}}
\caption{Same as Fig. \ref{fig3} for elliptical galaxies.}
\label{fig4}
\end{figure}

\begin{figure} 
\centering
\mbox{\psfig{figure=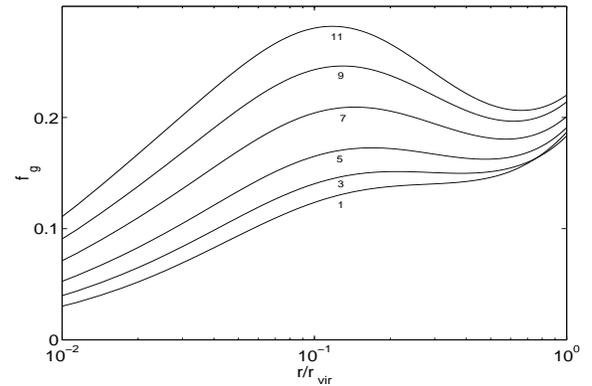,height=2.0in,width=3.0in}}
\caption{Gas mass fraction vs radius in clusters. The numbers on the curves correspond to the elapsed time in Gyr divided by $F_B\times (T/10^8 K)^{-3/2}$.}
\label{fig5}
\end{figure}

\begin{figure} 
\centering
\mbox{\psfig{figure=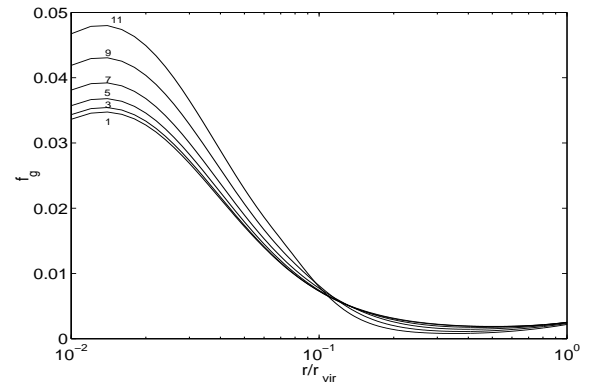,height=2.0in,width=3.0in}}
\caption{Same as Fig. \ref{fig5} for elliptical galaxies. }
\label{fig6}
\end{figure}

We solved equations (\ref{cont2})--(\ref{burg2}) and (\ref{newt}) for the
above initial conditions.
Because the gas was assumed to be isothermal, the evolution of metallicity
and helium abundance (fig. \ref{fig1}--\ref{fig4}) depends mainly on the
local gas mass fraction (fig. \ref{fig5} and \ref{fig6}). In clusters, the
gas mass fraction declines monotonically with radius initially, accounting
for the rise in metallicity and helium abundance towards the center. In
ellipticals, the gas fraction profile is more complex, which in turn
results in non-monotonic abundance profiles.

\section{Discussion}
Our calculations indicate that in the absence of magnetic suppression, the
diffusion timescale in hot clusters and cD galaxies is comparable to the
age of these systems. For a suppression factor $F_B\ga 5$, the diffusion
can be well described in the perturbative regime (\S 2.2) for most clusters
and galaxies, except in the central regions of very hot clusters. The
abundance change depends on the gas mass fraction, and characteristic
numbers were derived by adopting $f_{g}=0.17$ and $f_{g}=0.017$ for
clusters and cD galaxies, respectively.  Interior to the virial radius of
clusters with an age $t\sim 5$ Gyr, the helium and metal abundance would
increase by a factor of $1+0.3(T/10^8~{\rm K})^{1.5}/F_B$ and
$1+0.15(T/10^8~{\rm K})^{1.5}/F_B$, respectively. For elliptical galaxies
with an age $t\sim 10$ Gyr, the corresponding factors are
$1+0.2(T/10^7~{\rm K})^{1.5}/F_B$ and $1+0.1(T/10^7~{\rm K})^{1.5}/F_B$,
respectively.  Although the effect of diffusion on the distribution of
metals is rather modest, its effect on the distribution of helium is more
substantial.  As indicated by Figures \ref{fig1} and \ref{fig2}, some
regions inside the hottest X-ray clusters or cD galaxies, show an increase
of $\sim 50\%$ of helium abundance for $F_B=3$.

Detection of the diffusion signatures can be used to gauge the effect of
magnetic fields which also limit thermal heat conduction from suppressing
cooling flows in the same environments.  Although the increased helium
abundance can not be observed directly, it leads to a number of indirect
observable signatures.

As shown by Fig. \ref{fig5}, diffusion results in a substantial steepening
of the gas distribution. This would in turn lead to an increase in the total
X-ray luminosity. Since diffusion is fastest in high temperature
clusters, this process may contribute to the discrepancy between the
observed luminosity-temperature relation $L\propto T^3$ (Mushotzky 1984;
Edge \& Stewart 1991; David et al. 1993) and the form $L\propto T^2$ expected
from self-similar arguments (Kaiser 1986).

The mass profile of clusters and elliptical galaxies is usually inferred
from the observed distribution of hot X-ray gas under the assumption of
hydrostatic equilibrium. Increasing the helium fraction by $\sim 50\%$ is
equivalent to increasing the mean molecular weight by $\sim 10\%$, which in
turn results in overestimating the mass also by $\sim 10\%$. Thus, in cases
where alternative methods for mass determination (such as gravitational
lensing or the Sunyaev-Zel'dovich effect) reach better precision, these
methods may be used to constrain the diffusion velocity and the magnetic
suppression coefficient.

The change in the helium fraction would also affect the evolution of stars
formed from the enriched gas. D'Antona et al. (2002), who analyzed globular
clusters, found that increasing the initial helium abundance by $20\%$
results in the formation of a very blue horizontal branch. It would be
particularly interesting to search observationally for the distinct
spectral and evolutionary characteristics of helium-rich stars within X-ray
clusters and cD galaxies.

\bigskip
\noindent
{\bf Acknowledgements}

We thank Adi Nusser for stimulating discussions and Marat Gilfanov for valuable comments.
This work was supported in part by NASA grant NAG 5-13292, by
NSF grants  AST-0071019, AST-0204514 (for A.L.) and by German Israeli Foundation for Scientific Research and Development.

\end{document}